\def\year{2022}\relax
\documentclass[letterpaper]{article} 
\usepackage{aaai22}  
\usepackage{times}  
\usepackage{helvet}  
\usepackage{courier}  
\usepackage[hyphens]{url}  
\usepackage{graphicx} 
\usepackage{natbib}  
\usepackage{caption} 
\DeclareCaptionStyle{ruled}{labelfont=normalfont,labelsep=colon,strut=off} 
\pdfoutput=1
\usepackage{algorithm}
\usepackage{algorithmic}
\usepackage{newfloat}
\usepackage{listings}
\floatstyle{ruled}
\newfloat{listing}{tb}{lst}{}
\floatname{listing}{Listing}
\usepackage{booktabs}
\usepackage{tabularx}
\usepackage{subcaption}
\usepackage{pifont}

\usepackage{colortbl}

\usepackage{tikz-dependency}
\usepackage[T1]{fontenc}
\usetikzlibrary{fit}
\usetikzlibrary{positioning}

\pgfdeclarelayer{bg}    
\pgfsetlayers{bg,main}  
\definecolor{plot1}{RGB}{243,96,157}
\definecolor{plot2}{RGB}{157,210,255}
\definecolor{plot3}{RGB}{125,167,125}
\definecolor{plot4}{RGB}{247,147,29}
\definecolor{plot5}{RGB}{230,130,50}
\definecolor{plot6}{RGB}{130,67,34}
\definecolor{plot7}{RGB}{186,187,104}
\definecolor{plot8}{RGB}{30,67,34}
\definecolor{plot9}{RGB}{98,127,220}
\definecolor{plot10}{RGB}{247,47,29}
\definecolor{decentgrey}{RGB}{232,232,252}
\definecolor{mongo}{RGB}{220,170,51}
\definecolor{royal}{RGB}{0,68,136}

\newcounter{textboxno}
\usepackage[most]{tcolorbox}
\newtcolorbox{postbox}[3][]{
boxsep=0.05cm,left=0.05cm,right=0.05cm,bottom=0.001pt,
width=\columnwidth,
boxrule=0.03cm,
colbacktitle=white,coltitle=black,
boxed title style={
font size = \small,colframe=white,boxrule=0pt}, 
interior style={white},
enhanced,
float,
fonttitle=\scshape,
title={\refstepcounter{textboxno}\label{#1} Example \arabic{textboxno}: {#2}}
}

\usepackage{float}

\usepackage{tikz}
\usepackage{standalone}

\usepackage{pgfplots}
\pgfplotsset{compat=newest}
\usepgfplotslibrary{groupplots}
\usetikzlibrary{pgfplots.statistics}

\usepackage{xparse}   
\NewDocumentCommand{\rot}{O{90} O{1em} m}{\makebox[#2][l]{\rotatebox{#1}{#3}}}%

\usepackage{amsthm}
\usepackage{amsmath,amsfonts}

\newcommand{\fqt}[1]{``#1''}

\DeclareMathAlphabet{\mathsl}{OT1}{ptm}{m}{sl}

\definecolor{darkblue}{RGB}{0, 130, 185}
\definecolor{shallowblue}{RGB}{61, 223, 240}
\definecolor{darkyellow}{RGB}{183, 141, 36}
\definecolor{shallowyellow}{RGB}{252, 229, 80}
\definecolor{mygrey}{RGB}{204, 204, 204}

\definecolor{mongo}{RGB}{236, 142, 51}
\newcommand{\mongosquare}[1]
{{\protect\tikz\protect\draw[fill=mongo,draw=none] (0,0) rectangle ++(0.2,0.2); #1}}
\newcommand{\mongosquarel}[1]
{{\protect\tikz\protect\draw[fill=mongo!60,draw=none] (0,0) rectangle ++(0.2,0.2); #1}}
\newcommand{\mongosquarell}[1]
{{\protect\tikz\protect\draw[fill=mongo!30,draw=none] (0,0) rectangle ++(0.2,0.2); #1}}

\definecolor{royalblue}{RGB}{65,102,245}
\newcommand{\royalbluesquare}[1]{{\protect\tikz\protect\draw[fill=royalblue, draw=none] (0,0) rectangle ++(0.2,0.2); #1}}
\newcommand{\royalbluesquarel}[1]{{\protect\tikz\protect\draw[fill=royalblue!60, draw=none] (0,0) rectangle ++(0.2,0.2); #1}}
\newcommand{\royalbluesquarell}[1]{{\protect\tikz\protect\draw[fill=royalblue!30, draw=none] (0,0) rectangle ++(0.2,0.2); #1}}

\usepackage{xspace}

\newcommand{\citepos}[1]{\citeauthor{#1}'s \citeyearpar{#1}}

\newcommand{\ifsubmit}[1]{}

\newcommand{\be}{\begin{itemize}}
\newcommand{\ee}{\end{itemize}}
\newcommand{\bn}{\begin{enumerate}}
\newcommand{\en}{\end{enumerate}}
\newcommand{\bc}{\begin{center}}
\newcommand{\ec}{\end{center}}
\newcommand{\bl}{\begin{flushleft}}
\newcommand{\el}{\end{flushleft}}
\newcommand{\beq}{\begin{equation}}
\newcommand{\eeq}{\end{equation}}
\newcommand{\bq}{\begin{quote}}
\newcommand{\eq}{\end{quote}}

\newcommand{\bmp}{\begin{minipage}}
\newcommand{\emp}{\end{minipage}}

\DeclareMathAlphabet{\mathsl}{OT1}{ptm}{m}{sl}

\usepackage{xcolor}
\usepackage{siunitx}
\usepackage[np]{numprint}
\usepackage{multirow}
\npthousandsep{,}
\newcolumntype{T}{>{\tiny}l} 
\newcolumntype{H}{>{\Huge}l} 
\usepackage[inline]{enumitem}
\setlist[description]{leftmargin=1em}
\setlist[itemize]{leftmargin=1em}
\setlist[enumerate]{leftmargin=1.5em}

\usepackage{tikz}
\usepackage{standalone}
\usepackage{tikz-qtree}
\usepackage{float}

\usepackage[normalem]{ulem}

\title{Morality in The Mundane: \\ Categorizing Moral Reasoning in Real-life Social Situations
}
\author{
    Submission ID: 1241    
}

\begin{document}

\maketitle
\pagestyle{plain}

\begin{abstract}
Moral reasoning reflects how people acquire and apply moral rules in particular situations.
With social interactions increasingly happening online, social media data provides an unprecedented opportunity to assess \emph{in-the-wild} moral reasoning.
We investigate the commonsense aspects of morality empirically using data from a Reddit subcommunity (i.e., a subreddit) where an author may describe their behavior in a situation to seek comments about whether that behavior was appropriate.
A situation may decide other users' comments to provide \emph{judgments} and \emph{reasoning}.

We focus on the novel problem of understanding the moral reasoning implicit in user comments about the \emph{propriety of an author's behavior}.
Specifically, we explore associations between the common elements of the indicated reasoning and the extractable social factors.
Our results suggest that a moral response depends on the author's gender and the topic of a post.
Typical situations and behaviors include expressing \textsl{anger} emotion and using \textsl{sensible} words (e.g., f-ck, hell, and damn) in \textsl{work}-related situations.
Moreover, we find that commonly expressed reasons also depend on commenters' interests.
\end{abstract}

\section{Introduction}
\label{sec:introduction}
Moral reasoning concerns what people ought to do, which involves forming moral judgments in social or other situations \cite{sep-reasoning-moral}.
Researchers have extensively studied moral reasoning for investigating moral developments in groups organized by elements of social identity, based on genders \cite{bussey-1982-gender}, age \cite{lawrence-1989-long}, and profession \cite{wood-1988-ethical}.
These laboratory experiments are primarily conducted using questionnaires and hypothetical social situations that make the conflicts between moral principles stark.
However, real-life situations are nuanced and complex, and often present a wide variety of comparatively low-stakes decisions.
Social media provide an opportunity to assess the perception of normal social situations, such as understanding others' decisions on (im)morality of behaviors \cite{lourie-2020-scruples}.

\begin{figure}
    \centering
    \includegraphics[clip, trim=0.2cm 8cm 15cm 4cm, scale=0.6]{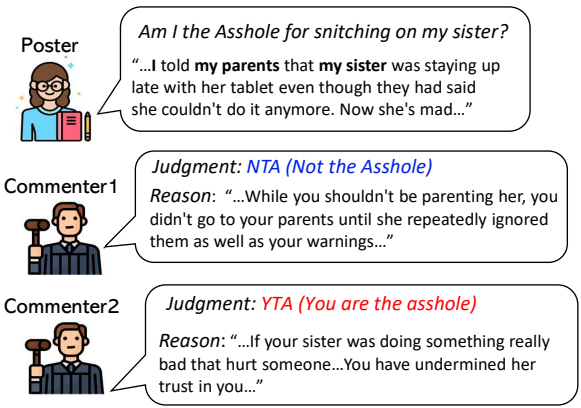}
    \caption{Sample post with comments where the final verdict (Not the Asshole) is decided by majority vote from the commenters. The post involves three parties - \emph{I}, \emph{my parents}, and \emph{my sister}. Commenters provide judgments and reasons about whether the author's behavior was inappropriate.}
    \label{fig:teaser}
\end{figure}

In this work, we study in-the-wild moral reasoning by examining a popular subcommunity of Reddit (i.e., subreddit) called /r/AmITheAsshole (AITA).\footnote{\url{https://www.reddit.com/r/AmItheAsshole/}}
In AITA, a user (i.e., \emph{author}) post interpersonal conflicts seeking others' opinions on whether their behaviors were appropriate.
AITA defines a few verdict codes, such as \textcolor{blue}{NTA} indicate authors' behaviors are appropriate, whereas \textcolor{red}{YTA} indicate authors' behaviors are inappropriate.
Other community members (i.e., \emph{commenters}) may comment on a post to provide moral \emph{judgments} (i.e., verdicts, justifying the verdict (if any) and other moral assessment) and the \emph{reasoning}.
Figure~\ref{fig:teaser} shows a post along with comments on it.
\begin{figure*}[!htb]
    \centering
    \includegraphics[clip, trim=0cm 12.5cm 0.5cm 6.4cm, width=\textwidth]{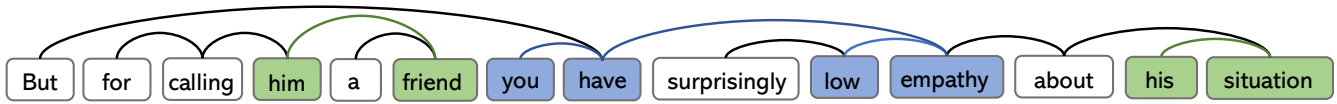}
    \caption{Dependency graph representation of an example comment. The shading shows syntactic relations.}
    \label{fig:example}
\end{figure*}
\begin{figure}[!htb]
    \centering
    \includegraphics[clip, trim=0.1cm 2.8cm 6.5cm 3.2cm, width=1\columnwidth]{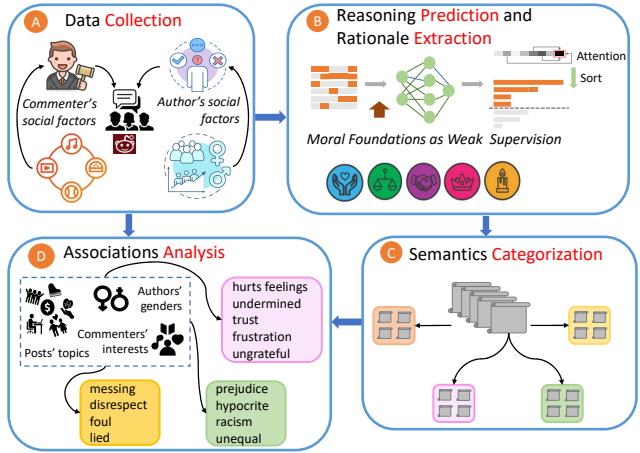}
    \caption{Flowchart depicting our research pipeline.}
    \label{fig:framework}
\end{figure}
Each comment includes a predefined community code along with an explanation for it.
A verdict of a post is decided by the top-voted comment's verdict.
Recent works focus on predicting verdicts of the posts and comments \cite{lourie-2020-scruples, zhou-2021-assessing, botzer-2022-analysis, xi-2023-blame}.
Another line of work analyzes the community using statistical methods \cite{botzer-2022-analysis, nguyen-2022-mapping, candia-2022-demo}.

However, to the best of our knowledge, no empirical work has conducted a systematic analysis to understand the reasoning implicit in the comments.
This paper focuses on the commonsense aspects of moral reasoning.
We apply Natural Language Processing (NLP) tools to investigate how the authors' and commenters' social factors shape their distributions and affect moral reasoning.
We extract authors' social factors from the posts.
These include \emph{authors' self-reported genders} and \emph{posts' topics}.
Here, we regard a post topic as a part of the author's social factors because the topic provides social information about the author, such as whether the author has had conflicts in marriage. 
As in previous work, we focus on authors' self-reported genders, not the genders of others involved \cite{botzer-2022-analysis, candia-2022-demo}.
For commenters' social factors as proxies of their \emph{interests}, we leverage the subreddits in which they participate \cite{candia-2022-demo}.
Extracting social factors from social media submissions has been extensively studied from various viewpoints, such as language bias \cite{ferrer-2021-biasreddit} and contentious conversations \cite{beel-2022-linguistic}. 

The contextual content in the reasoning can determine the verdict.
For instance, in Figure~\ref{fig:example}, the phrase \emph{low empathy} refers to the author's behavior and determines a \textcolor{red}{YTA} verdict. 
With a large corpus, these verdict-determining factors (i.e., \textbf{rationales}) would accumulate and reveal the common elements implicit in reasoning about specific social situations.
Therefore, we reformulate our task as building a computational \emph{predict-then-extract} model for categorizing the common elements of the moral reasoning embedded in the comments.
Figure~\ref{fig:framework} describes our research pipeline. 
Our proposed method involves predicting reasoning and extracting the rationales, categorizing them by meaning similarities, and analyzing their associations with the abovementioned factors.

\begin{description}
\item[Prediction] As discussed above, we distinguish the rationales that refer to authors from those that refer to others. 
Therefore, we consider the meaning and syntactic features (as shown in Figure~\ref{fig:example}) in references to various parties.
Specifically, we build a dual-channel context-feature extractor to obtain the global and local context features of sentences in the original post.
We evaluate our method in terms of its prediction performance.

\item[Extraction] We apply the rationalization process \cite{lei-2016-rationalizing, bastings-2019-interpretable, deyoung-2020-eraser} to extract rationales from the reasoning.
The selected rationales are small but sufficient parts of the input texts that \textsl{accurately} \cite{jain-2020-learning} identify the most important information actually used by a neural model.
Unlike previous works, we assume no human annotated labels for rationales on social media data.
Therefore, we follow \citet{jiang-2021-structurizing} to \fqt{weakly} label rationales using a domain-related lexicon, the Moral Foundation Theory (MFT) \cite{haidt-2007-mft}.
We then evaluate multiple methods to select plausible rationales.

\item[Categorization and Analysis] We apply k-means clustering \cite{lloyd-1982-kmeans} on the embedding vectors of the rationales and categorize their meaning commonality using a meaning analysis system.
Finally, we perform fine-grained analysis on the resulting meaning clusters. 
\end{description}

\paragraph{Findings and Contributions}
To the best of our knowledge, this is the first study to explore moral reasoning in AITA.
Through 51,803 posts and 3,675,452 comments, we find meaning commonalities associated with the authors' and commenters' social factors.
For example, female authors attract moral judgments expressing \textsl{angry} and \textsl{egoism} in \textsl{work}-related scenarios, while \textsl{politics} and \textsl{sensible} (e.g., f-ck, hell, and damn) are less likely present in such judgments. In addition, in \textsl{safety}-related situations, comments about \textsl{judgment of appearance} are more prevalent for female authors, whereas \textsl{physical/mental} (e.g., racist, homophobic, and misogynistic) are less likely to appear in the judgments.
Moreover, commenters interested in the \textsl{art} and \textsl{music} subreddits (e.g., r/AccidentalRenaissance) express more emotions such as \textsl{worry, concern, and confident}, than those interested in \textsl{news and politics}.

Our proposed model shows a 3\% improvement in all averaged scores (F1, precision, and recall) over finetuned BERT in predicting verdicts of the reasoning.
Moreover, our experiments demonstrate that with additional domain knowledge improve a rationale's plausibility.
The results indicate that our framework is effective in automatically understanding multiparty online discourses. 
Our framework is applicable in categorizing dynamic and unpredictable online discourse.
For instance, the framework can be applied in automated tools, such as for moderating rule-violating comments.
We have released our data and supplementary material.\footnote{\url{https://zenodo.org/record/7850027#.ZEGCCnaZM2w}} 
We will release our code once the paper is published.

\section{Related Work}

\paragraph{Moral Reasoning in Social Psychology}

Moral reasoning has long been studied.
\citet{bussey-1982-gender} find that moral decisions by males are typically based on law-and-order reasoning, while those by females are made from an emotional perspective. 
\citet{lawrence-1989-long} observe that participants' discussions about moral situations show clear age developmental trends over a two-year period.
\citet{wood-1988-ethical} report that individualism and egoism have a stronger influence on the moral reasoning on business ethics by professionals than by students. 
However, these studies do not provide a comprehensive understanding of moral reasoning on social media.

\paragraph{Morality in Social Media}

Social media helps ground descriptive ethics.
\citet{zhou-2021-assessing} profile linguistic features and show that the use of the first-person passive voice in a post correlates with receiving a negative judgment.
\citet{nguyen-2022-mapping} give a taxonomy of the structure of moral discussions.
\citet{lourie-2020-scruples} predict (im)morality using social norms collected from AITA.
\citet{forbes-2020-social} extract Rules of Thumb (RoT) from moral judgments of one-liner scenarios.
\citet{emelin-2020-moralstories} study social reasoning by constructing a crowd-sourced dataset including moral actions, intentions, and consequences.
\citet{jiang-2021-delphi} predict moral judgments on one-line natural language snippets from a wider range of possibilities.
\citet{ziems-2022-moral} build conversational agents to understand morality in dialogue systems.

\paragraph{Genders, Topics, and User Factors}

Gender differences are often relevant.
\citet{choud-2017-mental} reveal significant differences between the mental health contents and topics shared by female and male users.
\citet{candia-2022-demo} find young and male authors are likelier to receive negative judgments in AITA and society-relevant posts are likelier to receive negative moral judgments than romance-relevant posts in AITA.
\citet{ferrer-2021-biasreddit} find Reddit post topics are gender-biased; for instance, \textsl{judgment of appearances}-related posts are associated with females while \textsl{power}-related posts are associated with males.
Collecting personal information by using users' submissions on online platforms is a common method to explore social media data, such as investigating conversation divisiveness through Reddit \cite{beel-2022-linguistic}.

\section{Data}
\label{sec:data}

Reddit discussion structure is of a tree rooted at an initial post; comments reply to the root or to other comments.
\paragraph{Definitions}
We adopt definitions from \citet{guimaraes-2021-xpost} to describe instances in our dataset.
\begin{description}
    \item[A post] refers to the starting point in a discussion.
    \item[A top-level comment] refers to a comment that directly replies to a post.
\end{description}
We focus on top-level comments because other comments in AITA may not include judgments and reasoning based on the posts.

\subsection{Collection of Posts and Comments}

We require a large-scared corpus with relevant posts and comments.
Previous datasets are either nonpublic \cite{zhou-2021-assessing, candia-2022-demo, botzer-2022-analysis} or insufficient for our purposes \cite{lourie-2020-scruples, nguyen-2022-mapping}.
Therefore, we collected our dataset using PushShift API\footnote{\url{https://github.com/pushshift/api}} and Reddit API.\footnote{\url{https://www.reddit.com/dev/api}}
We scraped over 351,067 posts and the corresponding 10.3M top-level comments from AITA, spanning from its founding in June 2013 to November 2021.
We collected these submissions by applying rule-based filters following the aforementioned previous works to ensure their relevance and avoid discrepancies between data from Reddit and archived data from PushShift. 
We excluded deleted posts and comments because they may violate AITA rules, such as including fake content to solicit outrage.
We also exclude posts and comments submitted by deleted accounts and moderators.
We selected posts that have at least ten top-level comments to ensure quality.
We selected top-level comments that have a predefined code indicating the judgment and fifteen or more characters representing the reasoning.
Reddit allows users to give positive and negative feedback to submissions in the form of \textsl{upvotes} and \textsl{downvotes}.
Therefore, posts and comments in our dataset are associated with a \textsl{score}\footnote{Score is an aggregate of number reported by Reddit: \url{https://www.reddit.com/wiki/faq/#wiki_how_is_a_submission.27s_score_determined.3F}} representing the accumulated differences between upvotes and downvotes.

\paragraph{Extraction of Comments' Verdicts } 
The judgments are predefined codes: YTA (author's behavior is inappropriate), NTA (author's behavior is appropriate), ESH (everyone's behaviors are inappropriate), NAH (everyone's behaviors are appropriate), and INFO (more information needed).
Some comments use short phrases as codes (e.g., not the a-hole instead of NTA).
Therefore, we applied regular expressions to match such variants.
We resolved multiple matches by selecting the second match when there is a transition word such as \textsl{but}.
And, we reversed the extracted codes in judgments containing negations such as \textsl{I do not think} using regular expression. 
We removed sentences marked with $>$, which indicates a quotation. 
To evaluate the labeling process, we checked a random sample of 500 submissions. 
We found 5\% false positives and 6\% false negatives. 
Following \citet{lourie-2020-scruples}, we assigned labels to comments with YTA as 1, NTA as 0, and discard all other instances.

\subsection{Comment Corpus}

Our corpus selection criteria require that selected comments: (1) have scores higher than 100, (2) have a token length between 20 and 200, (3) have commenters who were previously awarded by a \emph{flair} (i.e., to select comments submitted by reputed users), and (4) replied to posts that contain authors' self-reported genders.
A flair is awarded by AITA and represents how many times a user's judgments have become the most upvoted comments, thus, reflecting the commenter's reputation.
As a result, our corpus includes 51,803 posts and 120,760 out of 3,675,452 total comments that belong to the selected posts.
The label distribution of NTA to YTA is 60--40.
We randomly selected 45,505 instances labeled as 0 and all instances (i.e., 45,505) labeled as 1.
We split our corpus as 80/10/10 for training, development, and testing.
Table~\ref{tab:data_sum} summarizes our dataset.
\begin{table}[htb]
    \centering
    \small
    \begin{tabular}{l r r r r}
    \toprule
    {} &Total& NTA & YTA&Mean \# Words\\
    \midrule
    Training & 72,808 & 36,405 &36,405 &184\\
    Development & 9,101 & 4,550 &4,550 &162\\ 
    Testing & 9,101 & 4,550 &4,550 &178\\
    \bottomrule
    \end{tabular}
    \caption{Dataset summary.}
    \label{tab:data_sum}
\end{table}

\section{Method}

This section introduces the processes of extracting \emph{social factors}, \emph{verdicts}, and \emph{rationales}.
There are two advantages to use rationalization for summarising common patterns of moral reasoning: (1) it can be trained with neural networks in an unsupervised manner \cite{deyoung-2020-eraser},\footnote{Our social media data is inherently without human annotated rationales as in previous works \cite{jain-2019-attention, jain-2020-learning, atanasova-2020-diagnostic}.} and (2) it provides appropriate rationales for social media data \cite{jiang-2021-structurizing}.

\subsection{Extraction of Topics, Genders, and Interests}

We adopt \citepos{nguyen-2022-mapping} topic model (with topics named by experts) to identify topics for posts in our corpus.
Then, we use regular expressions to extract authors' self-reported genders.
We leverage commenters' participation on Reddit to proxy their interests.

\paragraph{Topic Modeling for Posts' Topics}

Latent Dirichlet Allocation (LDA) \cite{blei-2003-lda} is widely applied for clustering text.
\citet{nguyen-2022-mapping} find 47 named topics in AITA posts via LDA models.
These topics are associated with clusters of words sorted by the probability of belonging to that topic.
We found that our corpus and \citepos{nguyen-2022-mapping} corpus have 34,098 posts in common, and the rest 17,705 posts were submitted after April 2020 (the ending time of their dataset).
We follow their method to assign each post the topic that has the highest prior probability.

\paragraph{Authors' Genders}

Extracting demographics from social submissions using regular expressions is common in analyzing Reddit data, such as in exploring contentious conversations \cite{beel-2022-linguistic}.
Gender and age are not typically available on Reddit, allowing for anonymous posting.
Fortunately, the social media template for posting gender and age, e.g., [25f] (25-year-old female) enables us to use regular expressions to extract the information. 
Note that authors typically report the demographic information of multiple parties in the situation described, such as \textsl{I [25f] and my wife [25m]}.
Therefore, we extract authors' self-reported genders by filtering first-person pronouns (i.e., I).
Besides, we consider gendered alternatives where available; for example, male can be estimated by \texttt{$\backslash$b(boy|father|son)$\backslash$b}) and female by (\texttt{$\backslash$b(girl|mother|daughter)$\backslash$b}).
We do not match nonbinary genders because we do not have ground truth labels for nonbinary targets.
As a result, we find the female/male split of 90--10 in our dataset
We took a random sample of 300 submissions to evaluate the regular expression. 
We found that gender extracted using our regular expression matches the manually labeled one 94\% of the time.

\paragraph{Commenters' Interests}

Following \citet{candia-2022-demo}, we proxy commenters' interests via the subreddits they participated in by making at least one submission (i.e., post or comment) within a six-month period (three months before and after the comment timestamp) based on the timestamp of the comment found in our corpus.
We focus on the commenters because they have made quality judgments.
We chose a six-month window based on users' prolificity, as defined by \citet{beel-2022-linguistic}, considering a user prolific if they submit more than 25 times in their interested subreddit. 
We manually checked 100 users and found that a six-month period makes users more prolific than four, eight, and twelve months.
We discard deleted user accounts, which restricts our analysis to \np{46519} commenters with \np{104915} comments.
Unlike \citepos{candia-2022-demo} work, we map the collected subreddits following Reddit's predefined subreddit categories.\footnote{\url{https://www.reddit.com/r/ListOfSubreddits/wiki/listofsubreddits/}}
Commenters may have interests in various categories.
Therefore, we set their interest as their most frequently submitted subreddit.

\subsection{Predicting Verdicts then Extracting Rationales}

We now introduce the rationalization process, followed by how our predict-then-extract model operates.

\begin{figure}[!htb]
    \centering
    \includegraphics[clip, trim=2.5cm 4.3cm 1cm 10.0cm, scale=0.53]{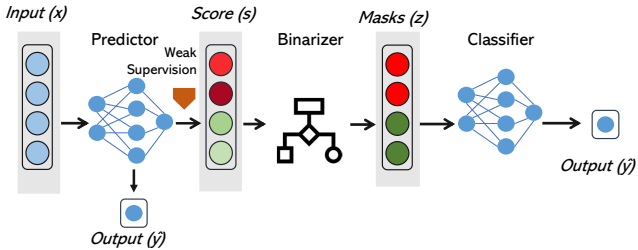}
    \caption{Soft rationalization is a three--phased process. The predictor outputs $\hat{y}$ and importance scores $s$. The binarizer assigns masks to tokens $z$. The classifier predicts unmasked tokens; it predicts $y$ again to evaluate a rationale's accuracy.}
    \label{fig:rationale}
\end{figure}
\begin{figure*}[!htb]
    \centering
    \includegraphics[clip, trim=1cm 7.5cm 3.5cm 5.5cm, scale=0.8]{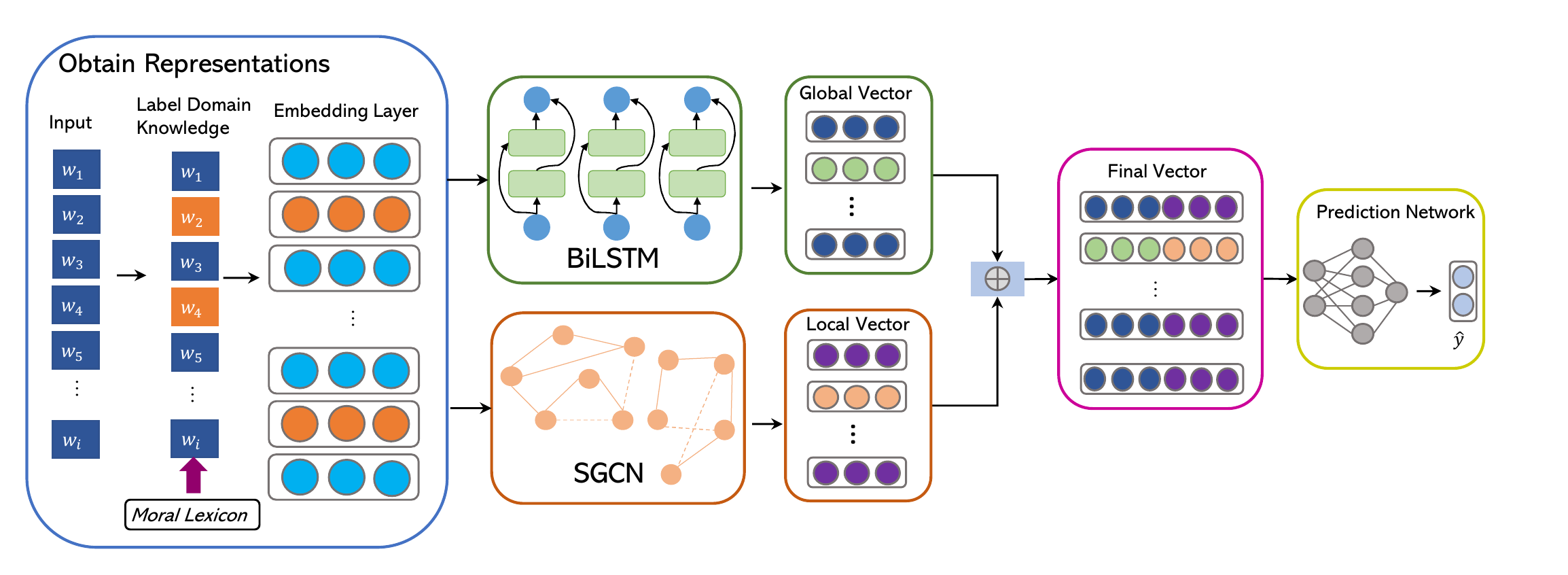}
    \caption{Architecture of the predictor in Figure~\ref{fig:rationale}. Here, $w_i$ represents a token in an input instance and $\widehat{y}$ is a predicted verdict. Tokens in \mongosquare are labeled by an additional moral lexicon.}
    \label{fig:architecture}
\end{figure*}

\paragraph{Introduction to Rationalization Process}

Given a pretrained model $\mathcal{M}$, each instance is of the form of $(x, y)$, where $x=[x^{i}]$ are the input tokens and $y\in\{0,1\}$ is the binary label.
The rationalization process outputs a predicted $\hat{y}$ with a binary mask $z=[z^{i}]\in\{0,1\}$ of input length, indicating which tokens are used to make the decision (i.e., $z^{i}=1$ if the $i$th token is used).
The tokens masked with ones are called rationales ($R$), and considered accurate explanations of the model's decisions and can be used alone to make correct predictions \cite{jain-2020-learning}.
Binarization methods are hard and soft according to \citep{deyoung-2020-eraser}.
Hard selection uses the Bernoulli distribution to sample binary masks (i.e., $z\sim$ Binarizer($x$)).
In contrast, soft selection \cite{jain-2020-learning} outputs multivariable distributions over tokens derived from features, e.g., self-attention values.
We adopt \textbf{soft selection} because hard selection faces performance limitations \cite{jain-2020-learning} and soft selection is more appropriate when there is no ground truth of rationales \cite{jiang-2021-structurizing}.

\paragraph{Prediction then Extraction}

Figure~\ref{fig:rationale} illustrates the architecture of a soft rationalization model.\footnote{Compared to \citet{lei-2016-rationalizing} we simplify the names of \textsl{encoder} as \textsl{predictor} and \textsl{generator} as \textsl{binarizer}. And we name \textsl{extractor} \cite{jain-2020-learning} as \textsl{binarizer}.}
The predictor in Figure~\ref{fig:rationale} is a standard text classification module that predicts a verdict.
We omit the last classifier module because we need the rationales instead of accurately predicting $y$.
The importance scores $z$ are computed via feature-scoring methods using the parameters (e.g., gradients) learned during training.
Therefore, the extracted rationales can capture the most salient contextual information used by a neural model when predicting a verdict.

Our experiments aim to empirically collect plausible rationales for categorizing the commonality of moral reasoning reflected in social media, instead of building an accurate predictor \cite{botzer-2022-analysis} or improving the rationalization extraction performance \cite{atanasova-2020-diagnostic, chrysostomou-2022-flexible}.

Figure~\ref{fig:architecture} shows our predictor.
We first weakly label tokens that appear in the moral lexicon.
We then obtain embeddings of input instances by adopting the pretrained \texttt{bert-base-uncased} model using Huggingface.\footnote{\url{https://huggingface.co/}}
Next, we prepare global and local representations of a sentence by a stacked Bidirectional LSTM (BiLSTM) \cite{hochreiter-1997-lstm} and a Syntactic Graph Convolutional Network (SGCN) \cite{bastings-2017-graph, li-2021-powering}.
Then, we feed the concatenated final hidden representation vectors into a fully connected prediction network.
The prediction network uses softmax to output the probabilities of a particular verdict.
We adopt cross-entropy in the network to measure loss. 

\textbf{Global context features} are multidimensional embeddings encoded using BERT \cite{Devlin-2019-BERT}, which maps a token into a vector based on its context.
We adopt the pretrained \texttt{bert-base-uncased} model from Huggingface to obtain embeddings. 
To obtain extended contexts, we use a Bidirectional LSTM (BiLSTM) \cite{hochreiter-1997-lstm}.
We compute the hidden states by passing the BERT-encoded embeddings to a stacked BiLSTM:
\begin{equation}
\overleftarrow{h_{g,i}}; \overrightarrow{h_{g,i}}=\text{BiLSTM}(S), i=1,2,
\label{eq:1}
\end{equation}
where $S$ represents the encoding output of the last layer of BERT and $i$ denotes the direction.
We compute the global context representations $h_{g,1}$ and $h_{g,2}$ by averaging the hidden outputs in both directions.

\textbf{Local context features} are obtained using a Syntactic Graph Convolutional Network (SGCN) \cite{bastings-2017-graph, li-2021-powering}, representing the local syntactic context of each token.
We capture words and phrases modifying the parties in input instances by using dependency graphs, which are obtained by applying the Stanford dependency parser \cite{chen-2014-fast} using Spacy.\footnote{\url{https://spacy.io/}}
The dependency graphs are composed of vertices (tokens) and directed edges (dependency relations), which capture the complex syntactic relationships between tokens.

SGCN operates on directed dependency graphs based on Graph Convolutional Network (GCN)
\cite{kipf-2016-semisupervised}.
GCN is a multilayer message propagation-based graph neural network.
Given a vertex $v$ in $G$ and its neighbors $\mathcal{N}(v)$, the vertex representation of $v$ on the $(j+1)$ layer is:
\begin{equation}
    h_v^{j+1} = \sum_{u\in\mathcal{N}(v)}W^{j}h_u^{j}+b^{j},
\end{equation}
where $W^{j}\in \mathbb{R}^{d^{j+1}\times d^{j}}$ and $b^{j}\in\mathbb{R}^{d^{j+1}}$ are trainable parameters, and $d^{j+1}$ and $d^{j}$ denote latent feature dimensions of the $(j+1)$ and the $j$ layers, respectively.
SGCN improves GCN by considering the directionality of edges, separating parameters for dependency labels, and applying edge-wise gating  \cite{bastings-2017-graph, li-2021-powering}.
Edge-wise gating can select impactful neighbors by controlling the gates for message propagation through edges.
Therefore, the SGCN module takes word embeddings and syntactic relations to compute local representations.
The local representation for a vertex (token) $v$ is:
\begin{equation}
h_{v}^{j+1}=\sum_{u\in N(v)}g_{u,v}^{j}(W_{d_{u,v}}^{j}h_{u}^{j}+b_{u,v}^{j}),
\label{eq:2}
\end{equation}
where $j$ represents a layer, $g$ is the gate on the $j$th layer to select impactful neighbors $u \in N$ of $v$, $W$ is the weight, and $b$ represents bias.
For each sentence, we use a pooling layer to convert tokens' local representations into a single hidden vector.

\textbf{Domain knowledge} is used to weakly label rationales, following \citep{jiang-2021-structurizing}. 
Such unsupervised rationalization favors informative tokens to optimize losses. 
However, our dataset's highly informative and frequent tokens such as gendered words (e.g., wife, boyfriend, and mother) may not determine the verdict.
Therefore, we use the popular \cite{nguyen-2022-mapping, ziems-2022-moral} psychological theory, Moral Foundation Theory (MFT) \cite{haidt-2007-mft}, to effectively select moral rationales.
MFT refines morality into five broad domains: care/harm, fairness/cheating, loyalty/betrayal, authority/subversion, and sanctity/degradation.
We adopt the extended version of the MFT lexicon \cite{Hopp2020TheEM}, containing \np{2041} unique words, to label comments in our corpus.
We reprocess the input instances and generate weak labels for rationales $z_{d}=[z_{d}^{i}]\in\{0,1\}$, where $z^{i}_{d}=1$ if $x^{i}$ is in the lexicon.
We include a loss term $L_{d}(z, z_{d})=-\sum_{i}|a^i|z^i_d$ for the soft selection process \cite{jiang-2021-structurizing}, where $a^i$ denotes the attention weight for token $z^{i}$. 
The term $L_{d}$ lowers the loss when the tokens selected by feature-scoring methods are morality-related; otherwise, it has no effect.
For prediction loss, we apply cross-entropy to optimize the network by calculating $L(y, \widehat{y})$ using the last hidden layer's output.
Combining the loss items, the objective of our model is:
\begin{equation}
    \arg \min L(y, \widehat{y})+\lambda L_{d}(z, z_{d}),
\end{equation}
where $\lambda$ controls the weight of domain knowledge loss.

\section{Experiments and Results}

We now evaluate of our predict-then-extract model.
For extraction performance, we first adopt multiple features scoring methods from previous works, followed by verifying the plausibility of the extracted rationales.

\subsection{Experimental Settings}
\label{sec:exp_settings}

\paragraph{Baseline Methods for Prediction}

We evaluate these machine learning models: the state-of-art transformer model BERT \cite{Devlin-2019-BERT}, Logistic Regression (LR), Random Forest (RF), and Support Vector Machine (SVM).
For LR, we use the lengths of the instances as a baseline model.
For traditional machine learning methods, we use GloVe \cite{pennington-2014-glove}, a text encoder that maps a word into a low-dimensional embedding vector, for textual classification to generate vector representations.
For BERT baseline, we apply the \fqt{Obtain Representations} and \fqt{Prediction Network} modules as shown in Figure~\ref{fig:architecture}.
We feed $\textrm{CLS}$ representations generated from the embedding layer into the prediction network instead of passing them through the BiLSTM and SGCN networks.

\paragraph{Feature Scoring Methods for Extraction}
We use random selection as a baseline and multiple feature-scoring methods to compute importance scores $s$:
\be{}
    \item Random (RAND): Randomly allocate importance scores.
    \item Attention ($\alpha$): Normalized attention weights \cite{jain-2020-learning}.
    \item Scaled Attention ($\alpha\nabla\alpha$): Attention weights multiplied by the corresponding gradients \cite{serrano-2019-attention}.
    \item Integrated Gradients (IG): The integral of the gradients from the baseline (zero embedding vector) to the original input \cite{sundararajan-2017-axiomatic}.
    \item Flexible (FLX): A flexible instance-level rationale selection method \cite{chrysostomou-2022-flexible}, under which each instance selects different scoring methods and lengths of rationales.
\ee{}
We compare only the above methods because they yield better performance than others, e.g., \cite{atanasova-2020-diagnostic}.

\paragraph{Evaluation Metrics for Rationales' Plausibility}
\label{sec:metric}
For prediction, we use macro F1-scores.
For extraction, we adopt metrics from previous works \cite{jain-2020-learning, chrysostomou-2022-flexible}:
\be{}
    \item reverse-Macro F1 (revF1): The performance of $\mathcal{M}$ in predicting $y$ when using full input and rationale-reduced input. The predicted label with full input is used as the gold standard. Masking rationales should drop the prediction performance; lower is better.
    \item Normalized Sufficiency (NS): 
    Reversed and normalized differences between predicting full input text and rationales: $max(0, 1-(p(\hat{y}|x)-p(\hat{y}|R)))$; higher is better. 
    \item Normalized Comprehensiveness (NC): Normalized differences between predicting full input text and rationale-reduced text: $p(\hat{y}|x)-p(\hat{y}|(x\not\in R))$; higher is better. 
\ee{}
Note that we are interested in generating plausible rationales, not producing accurate classifiers.
Therefore, we do not conduct human experiments to evaluate the accuracy of the generated rationales but to evaluate their plausibility \cite{chrysostomou-2022-flexible}, which in practice does not correlate with accuracy \cite{atanasova-2020-diagnostic}.

\paragraph{Hyperparameters}

For generating global representations, we use Adam optimization with an initial learning rate of $2\mathrm{e}{-5}$, $\epsilon=1\mathrm{e}{-8}$, a batch size of 16, \num{500} training steps, and a maximum sequence length of \num{256}.  
For generating local representations, the initial input to the first graph convolutional layer is the 768-dimensional global model representation.
These vectors are processed by the subsequent graph convolutional layer and output \num{128}-dimensional vectors.
The pooling layer for a vertex in Equation~\ref{eq:2} is a dense linear layer with tanh activation, whose input vectors are stacked vectors of all vertices and output is a single \num{128}-dimensional vector. 
We concatenate the global and local representations and obtain 896-dimensional vectors to feed into a prediction network.
The prediction network is a three-layer, fully connected, dense neural network, which comprises \num{512}, \num{256}, and \num{128} units, respectively, with ReLu activation.
To avoid overfitting, we regularize the prediction network using the Dropout technique; at each fully connected layer, we apply a Dropout level of $d = 0.5$.
Finally, the prediction network output is fed into the last neural network of two units; with softmax to obtain probability distributions of the verdicts.
We train five epochs for all the transformer-based models. 
All the experiments are implemented from Huggingface.

\subsection{Results}

The prediction performance of a model indicates its ability to distinguish commenters' evaluations of the various parties' behaviors.
The extraction performance of a model indicates the plausibility of the rationales it generates.

\paragraph{Performance of Predicting Verdicts}

\begin{table}[!htb]
\centering
\small
\begin{tabular}{l r r r}
\toprule
{Methods}& F1 (\%)& Precision (\%)& Recall (\%)\\
\midrule
    {LR-Length}&53.4&53.9&53.8\\
    {LR-GloVe}&57.0&57.7&56.2\\
    {Random Forest}&61.6&60.8&62.4\\
    {SVM}&63.8&63.2&65.3\\
    \midrule
    {BERT}&83.1&83.7&82.6\\ 
    {BERT-{Domain}}&82.6&82.8&82.5\\
    \midrule
    Global&83.0&83.0&82.8\\
    {{Global}-{Domain}}&82.6&83.7&81.5\\ 
    \midrule
    Local&83.5&82.9&84.2\\
    {{Local}-{Domain}}&83.7&83.6&83.6\\
    \midrule
    Global-Local&86.2&85.6&\textbf{86.9}\\
    {{Global}-{Local}-{Domain}}&\textbf{86.4}&\textbf{86.8}&86.1\\ 
    \bottomrule
    \end{tabular}
\caption{Macro F1 (the F1-scores calculated based on precision and recall scores), Precision, and Recall on the test set. The best scores are shown in bold (highest). Global-Local improves BERT by an average of 3.1\% on the three scores. Although BERT with domain knowledge does not outperform its counterpart without domain knowledge, the Global-Local-Domain method demonstrates an average of 3\% improvement on all three scores compared to BERT.}
\label{tab:predict_results}
\end{table}

\begin{table*}[!htb]
    \centering
    \small
    \begin{tabular}{l l >{\columncolor{mongo!80}[3pt]}r >{\columncolor{mongo!50}[3pt]}r >{\columncolor{mongo!80}[3pt]}r >{\columncolor{royalblue!80}[3pt]}r >{\columncolor{royalblue!50}[3pt]}r >{\columncolor{royalblue!80}[3pt]}r >{\columncolor{mongo!80}[3pt]}r >{\columncolor{mongo!50}[3pt]}r >{\columncolor{mongo!80}[3pt]}r}
    \toprule
    &&\multicolumn{3}{c}{Global} & \multicolumn{3}{c}{Local}&\multicolumn{3}{c}{Global-Local}\\
    \cmidrule(lr){3-5}\cmidrule(lr){6-8}\cmidrule(lr){9-11}
    &Methods&{revF1} &NS&NC&revF1 &NS&NC&revF1 &NS&NC\\
    \midrule
    \multirow{4}{*}{Domain}
    {}&RAND&85.9&0.26&0.27&79.3&0.25&0.27&84.0&0.20&0.34\\
    {}&$\alpha$ &59.2&0.31&0.42&56.0&0.33&0.53&52.5&0.45&0.64\\
    {}&$\alpha\nabla\alpha$ &58.1&\textbf{0.47}&0.61&45.8&0.50&0.77&42.9&0.47&\uwave{\textbf{0.81}}\\
    {}&IG&66.8&0.31&0.54&65.2&0.35&0.54&65.9&0.37&0.50\\ {}&FLX&\textbf{42.3}&0.52&\textbf{0.72}&\textbf{41.3}&\uwave{\textbf{0.59}}&0.77&\uwave{\textbf{38.9}}&\textbf{0.50}&0.80\\
    \midrule
    \multirow{4}{*}{No Domain}
    &RAND&79.0&0.25&0.30&86.6&0.24&0.29&88.0&0.21&0.33\\
    {}&$\alpha$ &62.1&0.29&0.39&62.5&0.37&0.61&63.6&0.38&0.61\\
    {}&$\alpha\nabla\alpha$ &57.2&0.37&0.65&56.9&0.45&0.64&54.1&0.45&0.70\\
    {}&IG&68.2&0.32&0.53&63.9&0.30&0.53&62.2&0.28&0.45\\
    {}&FLX&44.6&0.44&0.69&42.6&0.46&\textbf{0.78}&41.3&0.49&0.77\\
    \bottomrule
    \end{tabular}
    \caption{The Normalized Sufficiency (NS) and Normalized Comprehensiveness (NC) scores range over $[0,1]$. Results with \fqt{Domain} are with the domain knowledge module when predicting verdicts, results with \fqt{No Domain} are without the module. The best scores (the revF1 are the lowest; the NS and NC are the highest) in each column are shown in bold. The under-waved numbers are the highest NS and NC scores and lowest revF1 score among the three metrics. The averaged performance scores ($\lambda=0.1$) on the testing and development sets are similar. Among the three scores for the five feature-scoring methods (total fifteen for each prediction model), the number of times the Domain beats the No Domain for Global is \textbf{10 out of 15}, for Local \textbf{9 out of 15}, and for Global-Local \textbf{13 out of 15}.}
    \label{tab:rational_results}
\end{table*}

We use five-fold stratified cross-validation for the aforementioned classifiers.
For the transformer-based models, we ran each model with five epochs.
The reported performance score averages are shown in Table~\ref{tab:predict_results}.
The scores with domain knowledge are calculated with $\lambda=0.1$, which yields the best performance.
We observe that neural models outperform traditional machine learning models.
The Global-Local-Domain method shows an average of 3\% improvement among all the scores compared to a finetuned BERT.
We are unable to compare our results with \citet{botzer-2022-analysis} because of different research purposes and lack of their dataset and experimental details.

\begin{figure*}[!htb]
\centering
\begin{subfigure}[b]{\textwidth}
    \centering
    \includegraphics[scale=0.14]{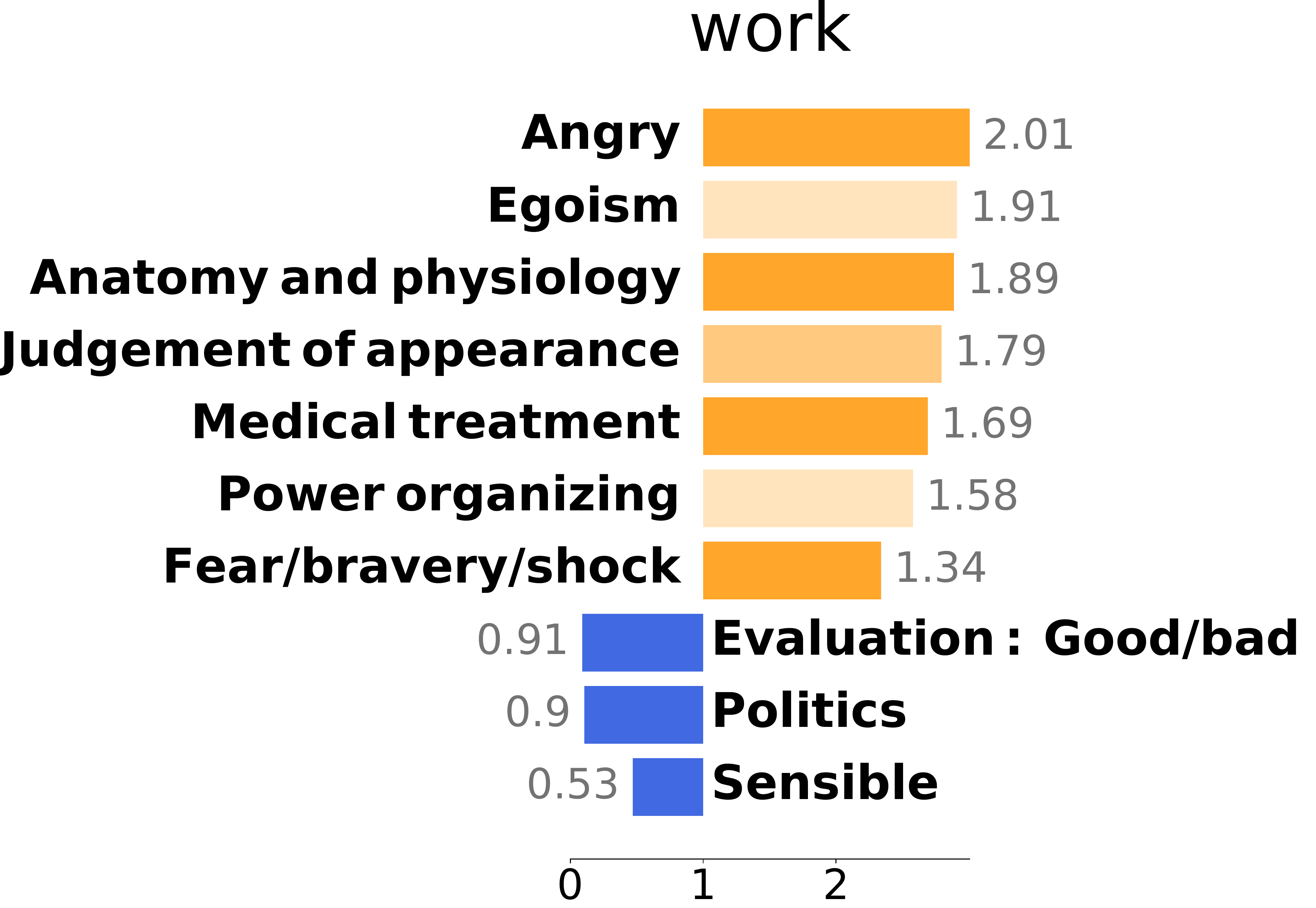}
    \includegraphics[scale=0.13]{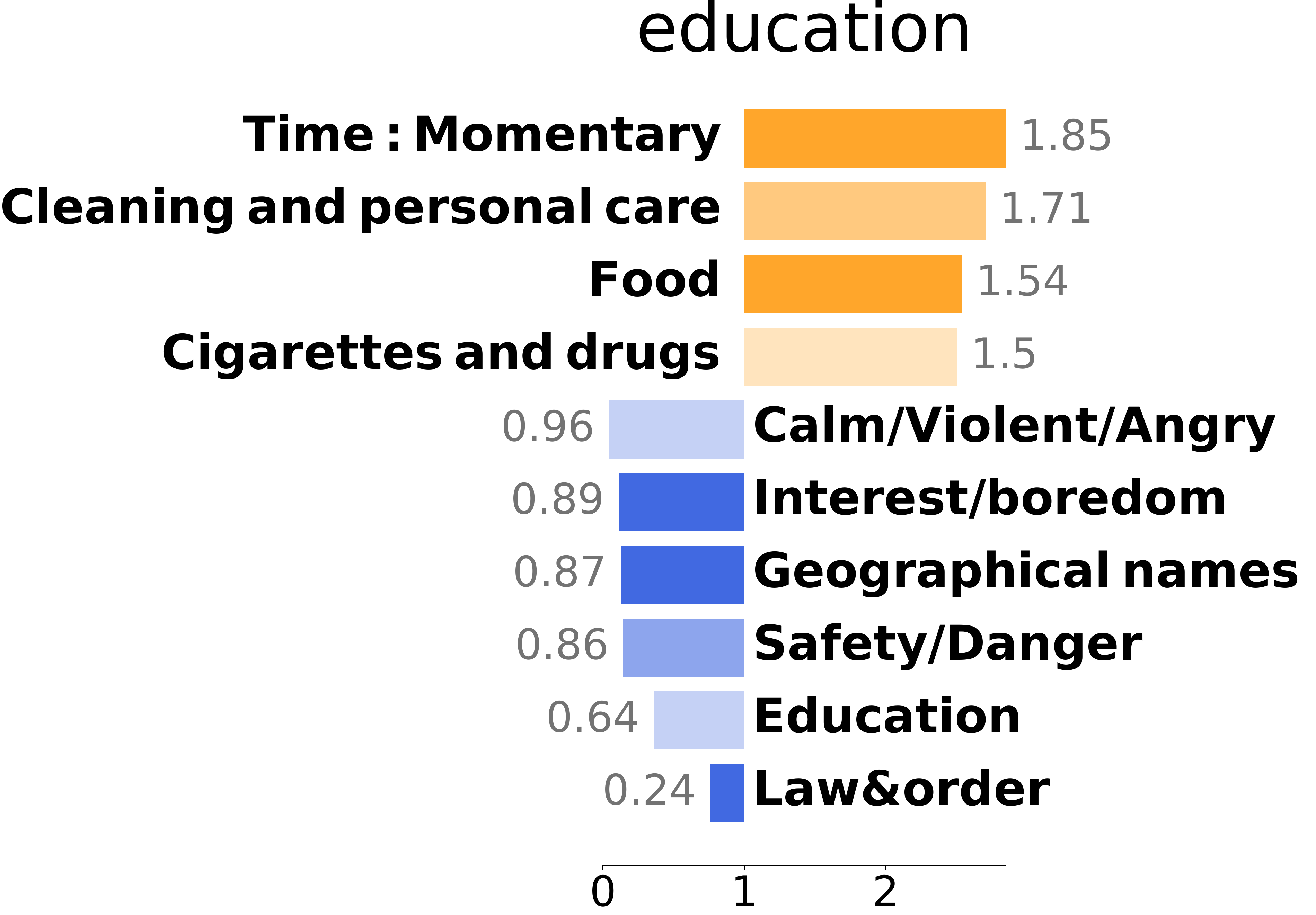}
    \includegraphics[scale=0.13]{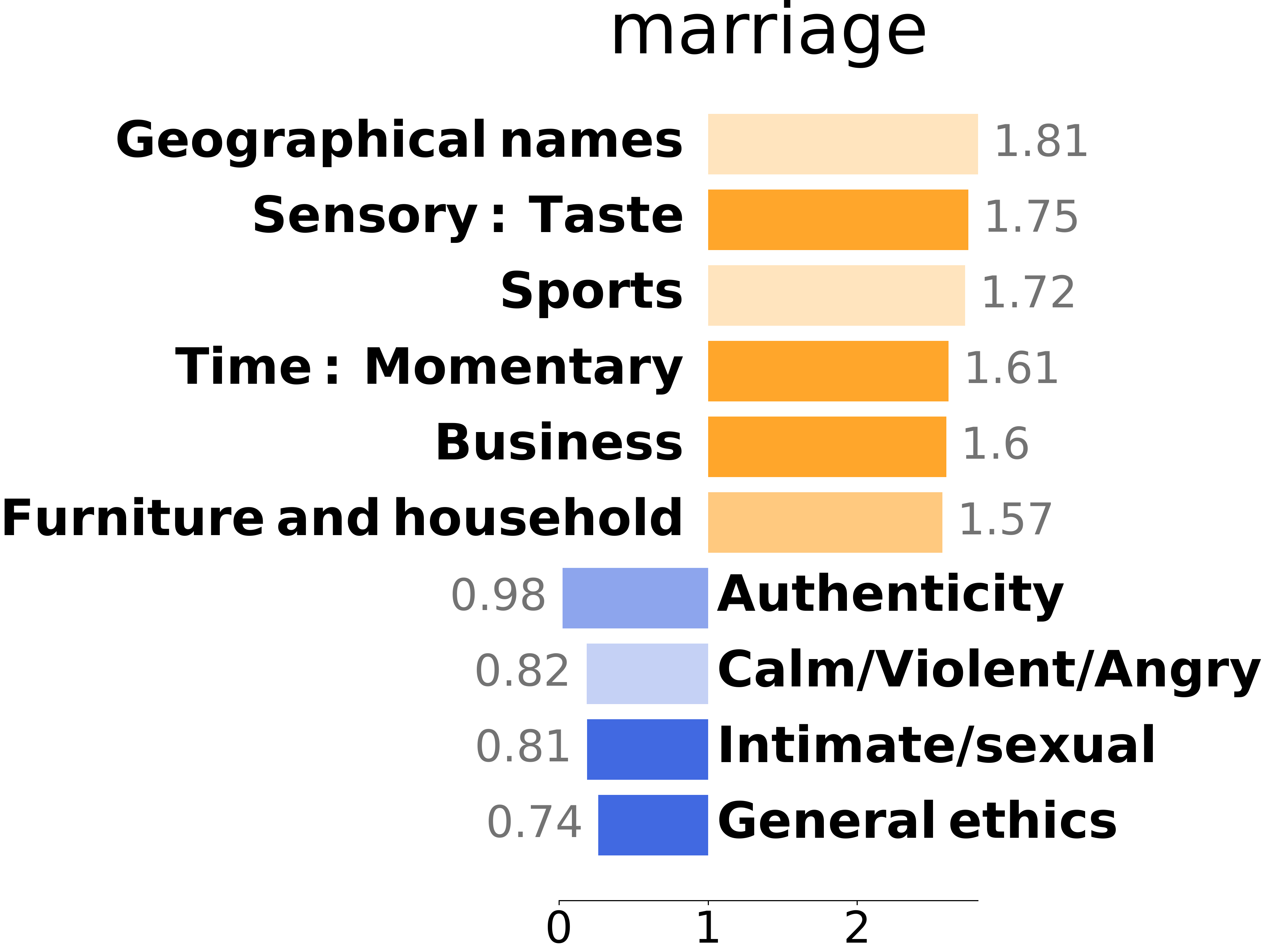}

    \smallskip
    \includegraphics[scale=0.13]{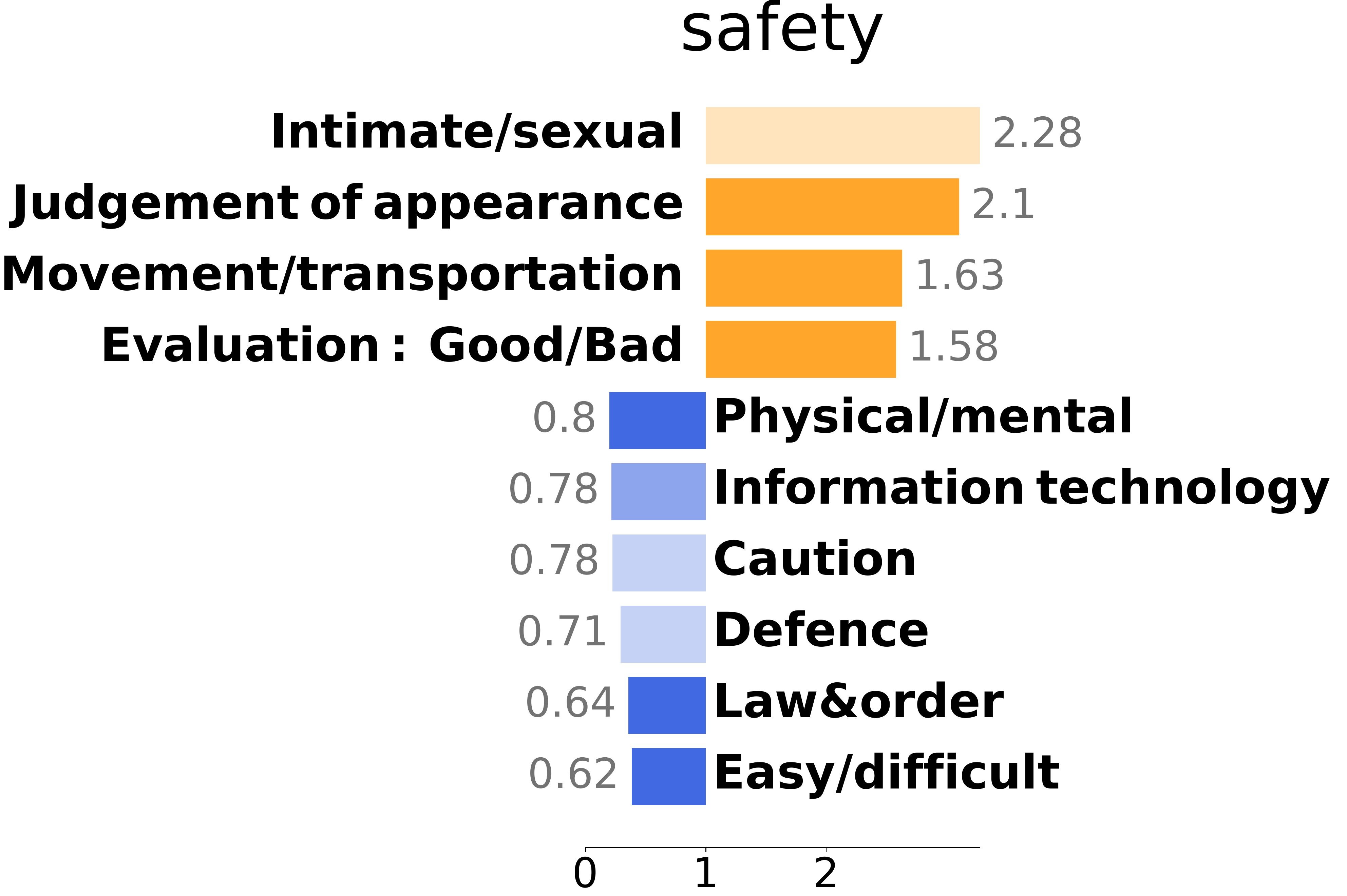} 
    \includegraphics[scale=0.13]{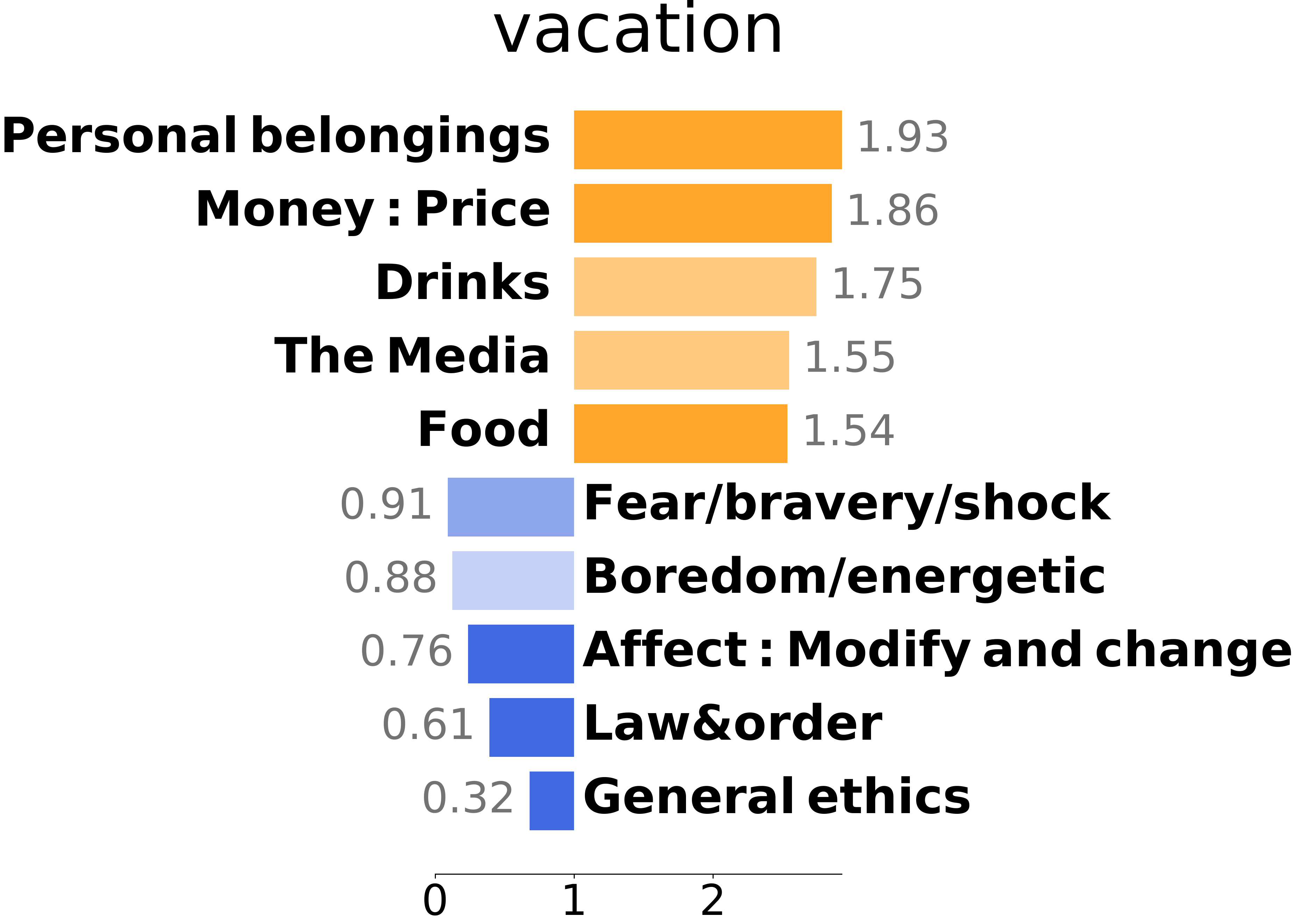} 
    \includegraphics[scale=0.13]{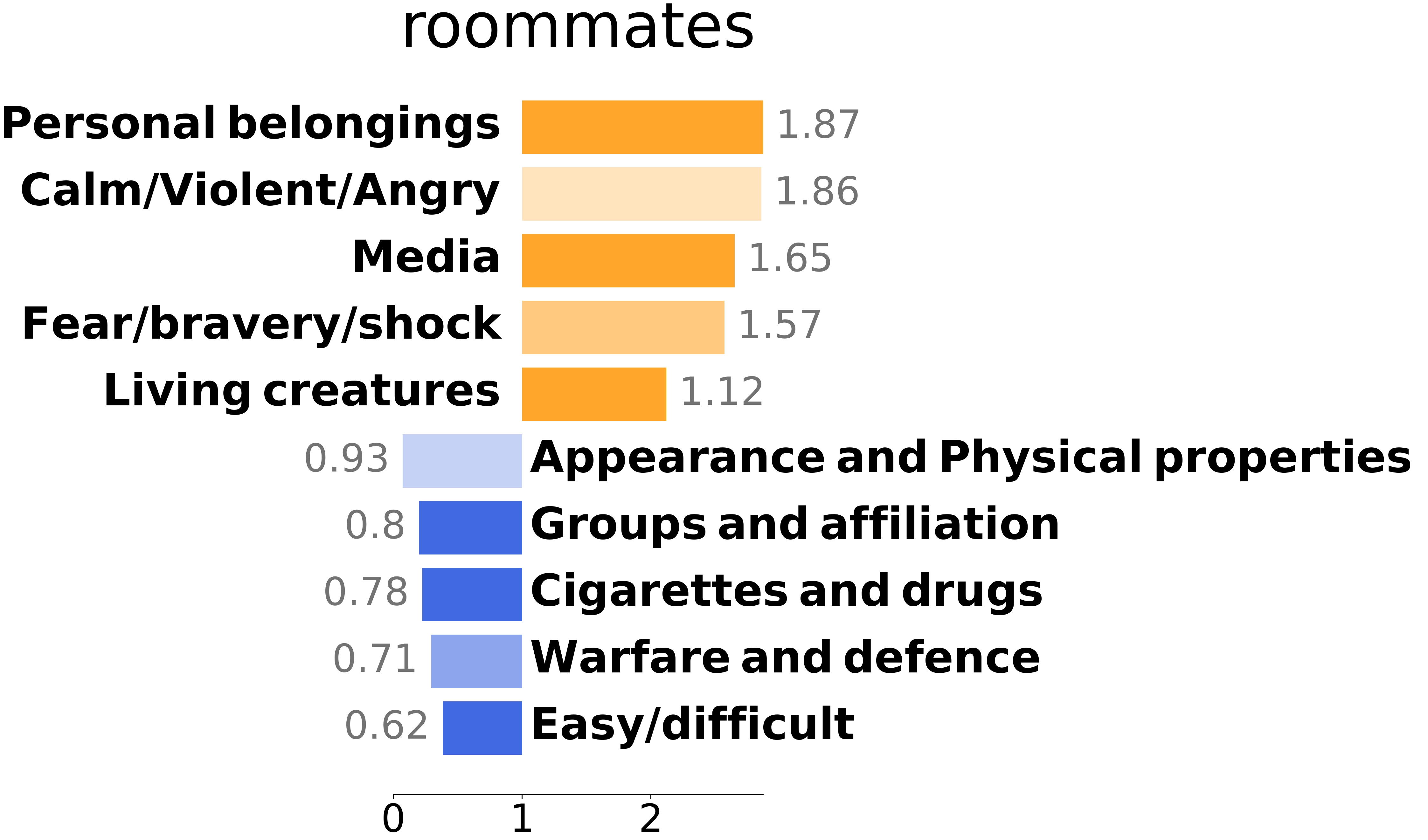} 
    \caption{The odds ratio values of authors' gender and meaning clusters in different topics. An odds ratio greater than one indicates the category is more likely to appears in the comments when the posters are females compared to males. And an odds ratio smaller than one indicates the opposite.}
    \label{fig:sub1}
    \end{subfigure}
\hfill%
\begin{subfigure}[b]{\textwidth}
    \centering
    \includegraphics[scale=0.38]{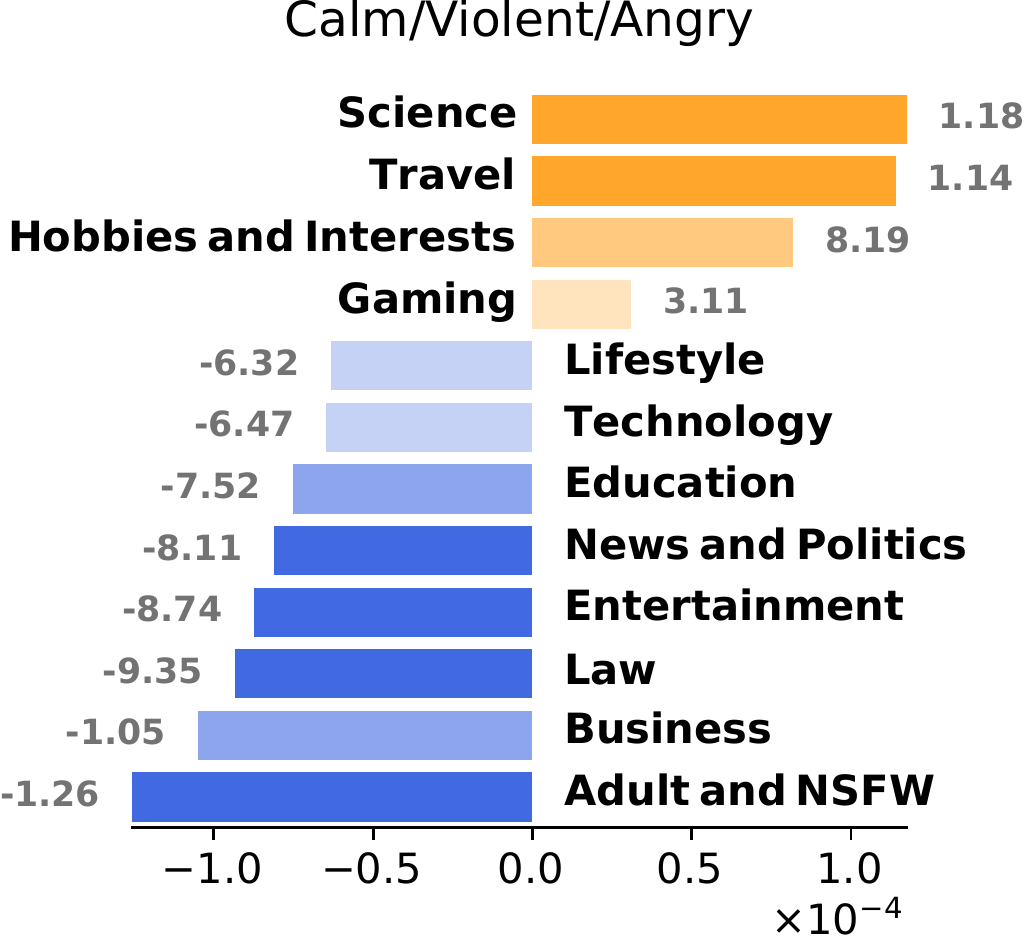}
    \includegraphics[scale=0.38]{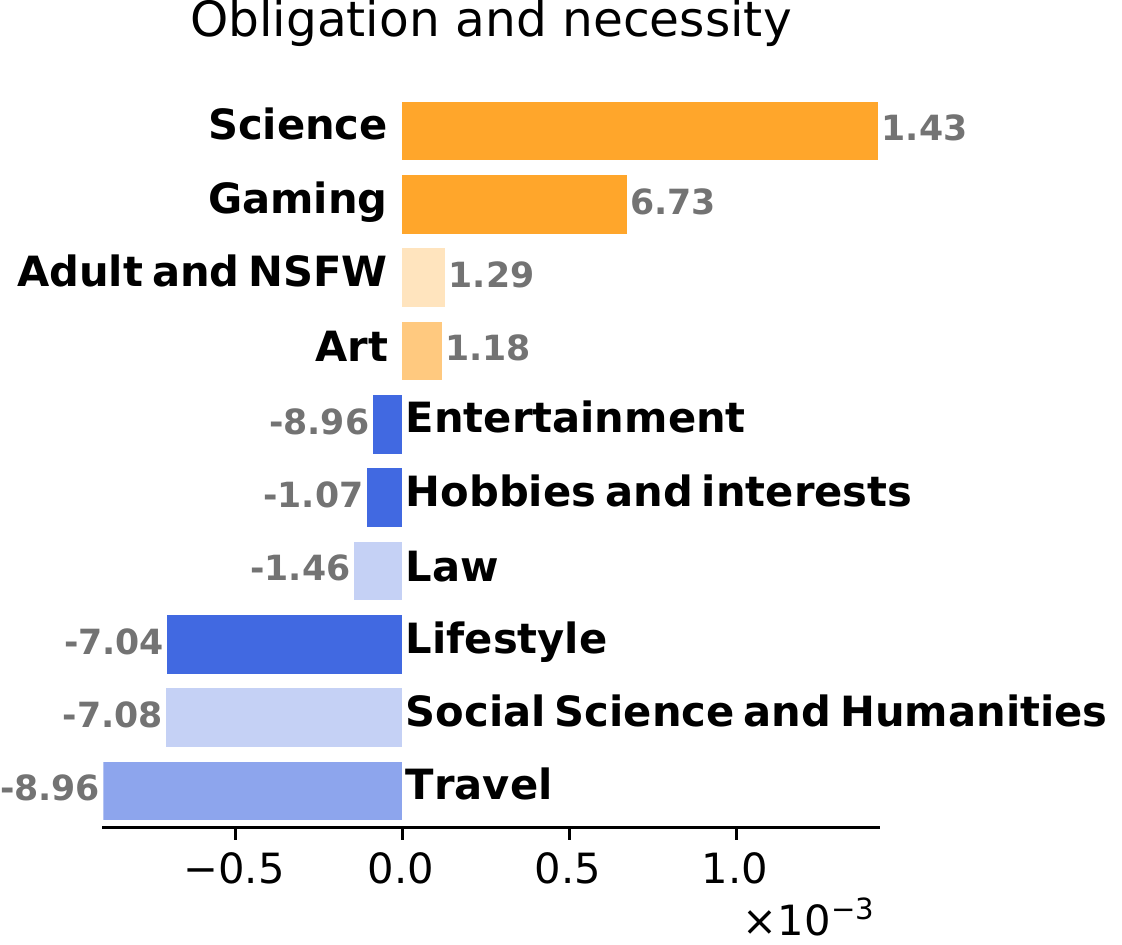}
    \includegraphics[scale=0.38]{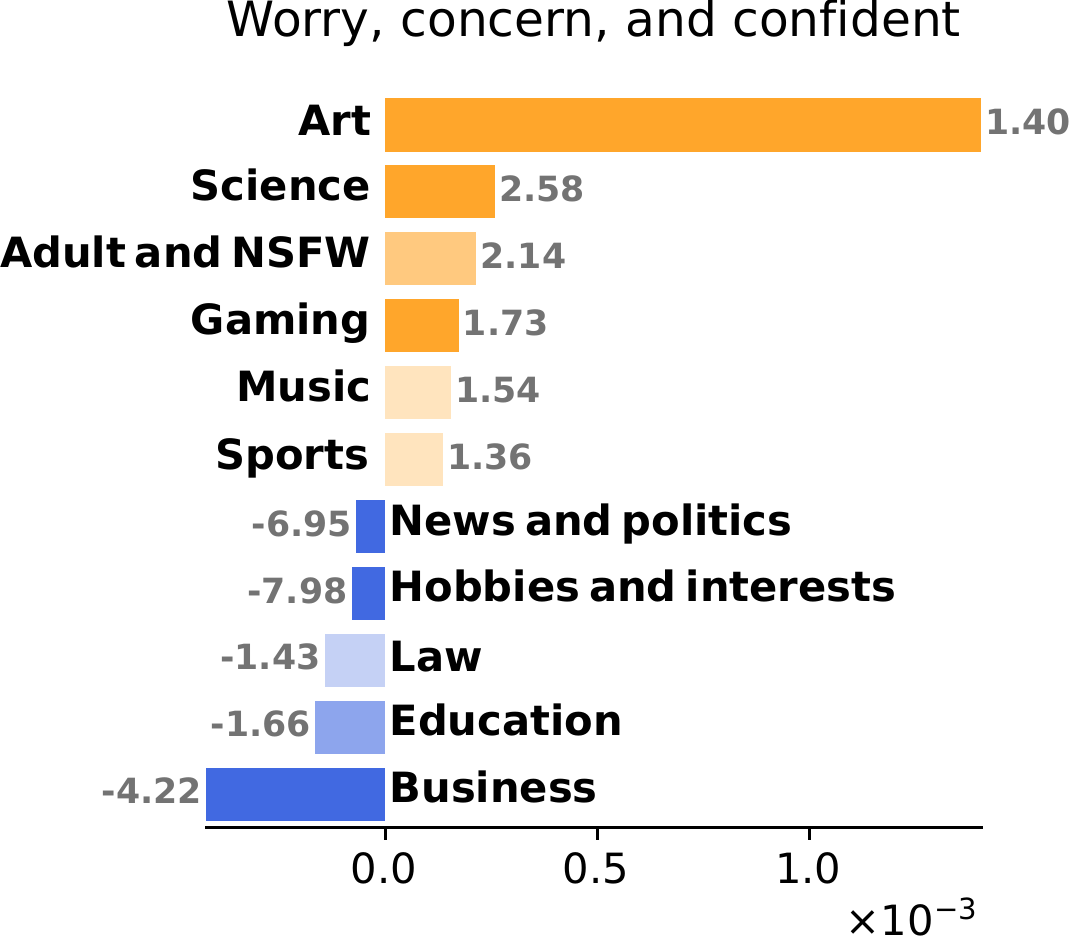}
    \includegraphics[scale=0.38]{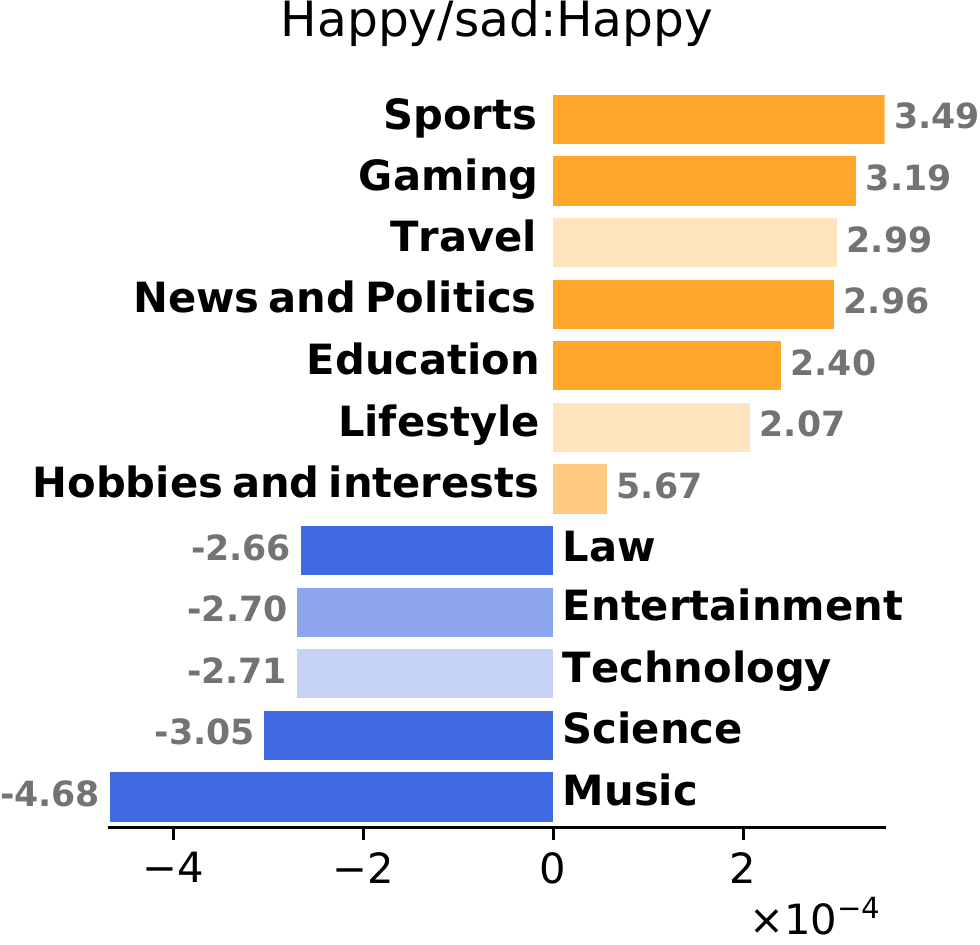}
    \caption{Regression results for the effects of the proxied commenters' interests. An effect that is greater than zero indicates positive effect. And an effect smaller than zero indicates the opposite.}
    \label{fig:sub2}
\end{subfigure}

\caption{We use orange rectangles \mongosquare{}to indicate odds ratio greater than one and effects greater than zero (on the right), and blue rectangles \royalbluesquare{}indicate the opposite (on the left). The shade shows the $p$-values: \mongosquare{}and \royalbluesquare{} (darkest): $\le0.0001$, \mongosquarel{}and \royalbluesquarel{} (middle): $\le0.001$, \mongosquarell{}and \royalbluesquarell{}: $\le0.05$. 
}
\label{fig:res_or}
\end{figure*}

\begin{table*}[!htb]
    \centering
    \small
    \begin{tabular}{l l l} 
    \toprule
    Clusters&Topics&Examples \\
    \midrule
    \multirow{2}{*}{\shortstack{Judgment of\\appearance}}& Work&skinny, curly, chubby, lean, eat, meat, slim, bodied, blonde, sickly \\
    {}& Safety&underwear, panties, bikini, clingy, thong, boudoir, swimsuit, bras, headband, earrings\\
    \midrule
    \multirow{2}{*}{\shortstack{Evaluation:\\ Good/Bad}}& Work&derogatory, \textsl{extremely terrible}, \textsl{horrible}, derisive, \textsl{awful}, \textsl{awesome}, pejorative \\
    {}& Safety&\textsl{awful}, \textsl{horrible}, \textsl{extremely terrible}, incredible, nightmare, \textsl{awesome}, amazing\\
    \midrule
    \multirow{3}{*}{Calm/Violent/Angry}&Marriage&kick, spitting, stomped, slapping, wasted, missed, fight, punching, cheating on, lied to\\
    {}&Education&picky, lived, mortified, resentful, nauseous, baffled, inconvenienced, conflicted\\
    \midrule
    \multirow{3}{*}{Law\&order}& Education&punish, punishment, jails, prison, inability, failure, lack, fault, mistakes, error\\
    {}& Safety&abuse, harassment, bullying, sexual, neglect, sexual, rape, abusers, cruelty, humiliation \\
    \midrule
    \multirow{2}{*}{Fear/bravery/shock}& Work&insecurities, humiliating, exhausting, miserable, sad, doubly, stressful, messy\\
    {}& Roommates&cruel, vile, despicable, cowardly, inhumane, atrocious, abhorrent, brutal, aggression\\
    \bottomrule
    \end{tabular}
    \caption{Examples of meaning clusters embedded in moral reasoning for topic-specific posts. Italics show the words that are common between the topics of the same cluster. The results indicate that words used in comments are different based on posts' topics. In addition, adverbs used in \textsl{Evaluation: Good/Bad} are more prevalent than adjectives used in \textsl{Judgment of Appearance}.}
    \label{tab:clusters}
\end{table*}

\paragraph{Ablation Studies for Prediction}
We perform ablation studies to understand how global and local representations affect prediction performance.
Table~\ref{tab:predict_results} shows that separately using global or local representations does not improve prediction performance over BERT, while combining both representations achieves the best performance.

\paragraph{Performance of Extracting Rationales}
Table~\ref{tab:rational_results} illustrates the performance of various feature-scoring methods with and without weakly supervision through domain knowledge.
We experiment on two scored token-selection methods \cite{jain-2020-learning}: (1) selecting the $K$ highest scoring (TopK) tokens for each instance and (2) selecting highest overall $K$-gram scoring tokens in the span of input tokens.
We adopt TopK for further analysis because it yields the best performance.
Although Table~\ref{tab:predict_results} shows that considering domain knowledge may not improve prediction performance for all neural models (e.g., BERT), we observe that using the instance-level rationale extraction method (FLX) with domain knowledge improves a rationale's plausibility.
Combining the results of Table~\ref{tab:predict_results} and Table~\ref{tab:rational_results}, we use Global-Local-Domain to predict and the FLX feature-scoring method to extract appropriate rationales.

\section{Analysis}
We leverage the \np{17808} identified rationales from the \np{18202} instances of our corpus's development and test sets as a lexicon.
We apply this lexicon on the total 3,675,452 comments in our corpus.
We introduce how we identify and cluster the extracted rationales' meanings.
Then, we perform a fine-grained analysis to investigate how the clusters are associated with the authors' and commenters' social factors.

\subsection{Clustering and Tagging} 

We apply pretrained GloVe embeddings \cite{pennington-2014-glove} to cluster the meaning similarities of rationales (i.e., averaged embeddings for phrases). 
After manually checking the clustered results, we select GloVe embeddings as they provide more detailed and informative clusters than other embedding methods, such as word2vec \cite{mikolov-2013-distribute}. 
We exclude the rationales that have negative dependencies in the original sentences to avoid ambiguity.
To aggregate the most similar embeddings into clusters, we employ the well-known k-means clustering algorithm. 
We tag the resulting clusters with USAS,\footnote{\url{http://ucrel.lancs.ac.uk/usas/}} a framework for automatic meaning analysis and tagging of text, which is based on McArthur's Longman Lexicon of Contemporary English \cite{mcarthur-1982-longman}. 
We use this lexicon \cite{piao-2015-development} to name the tags.
Because the generated rationales contain phrases and words, we filter the phrases composed of words belonging to the same USAS categories, such as \texttt{extremely awful}. 
We discard clusters of pronouns and prepositions. 
As a result, we find 86 unique meaning clusters in our dataset.

\subsection{Associations between Comments and Factors}

We measure the Odds Ratio (OR) to assess the associations between posts' topics, authors' genders, and meaning clusters.
For commenters' interests, we apply linear regression to compute their effects on the judgments. 
Figure~\ref{fig:res_or} and Table~\ref{tab:clusters} report the results.

\paragraph{Common but Distinct Reasoning in Topic-Specific Situations}

Our corpus includes six common post topics: work, education, safety, vacation, roommates, and marriage. 
Figure~\ref{fig:sub1} shows the topics with OR, indicating polarized comments for authors' genders.
Our analysis reveals consistent gender effects in certain categories, such as \textsl{Calm/Violent/Angry} in \textsl{education} and \textsl{marriage}, and \textsl{Judgment of appearance} in \textsl{work} and \textsl{safety}.
Moreover, as indicated in Table~\ref{tab:clusters}, we also observe distinct preferences for word usage to convey identical meanings even among categories with similar gender effects, suggesting nuanced and context-specific usage of language in moral reasoning.
However, the \textsl{Evaluation: Good/bad} category shows controversial effects, showing biases towards females in \textsl{safety} but towards males in \textsl{work}, indicating polarizing opinions.
Interestingly, the most contentious topics, such as relationships \cite{candia-2022-demo, ferrer-2021-biasreddit}, do not show typical gender biases in our analysis.
This could be due to the fact that commenters have specific evaluation standards in different moral scenarios. 

We find distinctive gender effects in \textsl{work} and \textsl{safety} (i.e., the maximum difference between OR scores is over \num{1.5}).
In such topics, words in \textsl{Sensible} and \textsl{Law\&order} are less likely used in comments towards female authors, and words related to \textsl{Judgement of appearance} are more likely to be.
The observation can be explained by the reflection of the persistent societal pressure on women to conform to certain beauty standards \cite{stuart-2012-choosing}. 
Moreover, commenters use different adjectives, verbs, and nouns to emphasize their concerns based on a given situation, while employing similar adverbs to express their emotions. 
For instance, the adjectives, verbs, and nouns used in \textsl{Judgment of appearance} for \textsl{work} and \textsl{safety} are dissimilar, whereas the adverbs employed in \textsl{Evaluation: Good/Bad} are common. 

\paragraph{Commenters' Interests Matter}

We now investigate how the commenters' interests (as proxied by the subreddits they participate in) affect their moral reasoning.
There are eighteen categories of subreddits that the commenters participated in (ordered in frequency): lifestyle, science, locations, technology, hobbies and interests, law, adult and NSFW, business, social science and humanities, music, sports, entertainment, news and politics, gaming, architecture, art, travel, and education.
These categories exhibit high popularity and diversity.
For example, \textsl{news and politics} includes subreddits, such as r/PoliticalHumor and r/antiwork, each with over a million users and \textsl{lifestyle} includes r/baking (over 1.6M members) and r/relationships (over 3.4M members).

The proxied commenters' interests are confounded with each other.
Therefore, we investigate them simultaneously to measure the causal effects of their interests.
We use an Ordinary Least Square (OLS), model\footnote{\url{https://www.statsmodels.org}} which is a common method for analyzing social variables \cite{ross-1980-causal}.
The following model captures the linear effects:
\begin{equation}
    b = \beta_0 + \beta_i x_i + \epsilon_i, i \le n,
\end{equation}
where $x_i$ denotes the frequency of the $i$th cluster appearing in judgments $b$, $\beta$ represents the constant effect of $x_i$, $n$ is the total number of clusters, and $\epsilon_i \sim \mathcal{N}(\mu,\,\sigma^{2})$ is normally distributed noise centered at 0.

Figure~\ref{fig:sub2} shows the effects of the interest categories on emotion-relevant clusters.
We observe that some categories such as \textsl{Sports} and \textsl{Lifestyle} are likelier to positively affect optimistic clusters \textsl{Happy/sad:Happy} than neutral clusters such as \textsl{Music}.
In addition, \textsl{Gaming} and \textsl{Science} positively affect using \textsl{Obligation and Necessity} words, such as \textsl{would}, \textsl{should}, and \textsl{must}.
Conversely, \textsl{Social Science and Humanities} and \textsl{Entertainment} have a negative effect.
The results may be explained by the distinctive personality traits of the social groups the commenters belong to.
For example, commenters interested in \textsl{Art} (e.g., r/AccidentalRenaissance) are the most likely to use \textsl{worry, concern, and confident} words and commenters interested in \textsl{Music} (e.g., r/NameThatSong) are the least likely to use  \textsl{Happy/sad:Happy} words.
A possible explanation may be that personalities of people interested in art are more emotionally sensitive than others \cite{csik-1973-personality}.

\section{Discussion and Conclusion}

Our research introduces a new framework for analyzing language on social media platforms. 
We focus on judgments of social situations and examine how social factors, such as a poster's gender and a commenter's interests, influence the distributions of common elements in the language used in comments. 
We employ NLP tools and a predict-then-extract model to collect these common elements.

Our study demonstrates that the language used in moral reasoning on AITA is influenced by users' social factors. 
For instance, consistent gender effects are observed in the \textsl{Calm/Violent/Angry} category in \textsl{education} and \textsl{marriage}, with posts authored by males more likely to receive such comments. Interestingly, our analysis reveals nuanced word usage within identical clusters, with verbs such as \fqt{kick} and \fqt{spitting} being frequently used in the \textsl{Calm/Violent/Angry} category in \textsl{marriage}, whereas adjectives such as \fqt{picky} and \fqt{mortified} were more common in \textsl{education}. 
Conversely, the \textsl{Evaluation: Good/bad} category in \textsl{work} and \textsl{safety} elicited conflicting opinions.

Our observations corroborate social psychology findings \cite{csik-1973-personality, stuart-2012-choosing}.
For example, comments about \textsl{Judgment of Appearance} in \textsl{work} and \textsl{safety} exhibit prevalence for female authors, which indicate the societal pressure on women to conform to beauty standards \cite{stuart-2012-choosing}.
Moreover, commenters interested in \textsl{Music} and \textsl{Art} are likelier to express emotions, which may be caused by their personalities \cite{csik-1973-personality}.
Overall, these findings highlight the context-specific and nuanced nature of language usage in moral reasoning on social media platforms, and contribute to a better understanding of the influence of social factors on language use.

\paragraph{Broader Perspectives}

Our research presents a novel framework for analyzing language usage in online media, which has practical implications for the design of monitoring systems to identify biased submissions in specific communities. 
This framework can assist commenters in reconsidering their comments and moderators in flagging concerning comments. 
In addition, the proposed methods can explain why a submission is considered biased and can inform the design of better features to educate new community members about problematic aspects of their submissions. 

Our findings align with social psychology research and shed light on societal pressures, such as the gendered pressure on appearance in work and safety contexts. 
Additionally, our study reveals the impact of personal interests on language use. 
These broader perspectives suggest potential implications for the development of more effective communication strategies online and underscore the need for further research exploring the relationship between language use and social factors in moral reasoning.

\paragraph{Limitations and Future Work}

Our empirical method inherently shares limitations with observational studies, e.g., susceptibility to bias and confounding.
There is a limit to how much we can tease apart social factors of the posters and commenters. 
We acknowledge some of the boundaries are unclear. 
For example, we treat genders as a social factors, but do the genders also affect posters' writing styles?
In addition to our single dataset analysis on AITA, there may be potential for further exploration on other data sets such as the \textsl{r/relationship\_advice} subreddit. 
Additionally, creating new datasets with crowd-sourced moral judgments could be beneficial in expanding the scope of analysis.

Although our prediction model takes into account the syntactic relations of input sentences, it is possible for some parties mentioned in a post to be background characters rather than active participants.
Moreover, the rationales we extract are not validated with ground-truth labels, mainly due to the complexity of the instances in our dataset.
In future work, we plan to leverage our framework to construct annotation guidelines to obtain human-evaluated clusters for analysis.

\paragraph{Ethics Statements}

Reddit is a prominent social media platform.
We scrape data from a subreddit using Reddit's publicly available official API and PushShift API, a widely used platform that ingests Reddit's official API data and collates the data into public data dumps.
None of the commenters' information was saved during our analysis.
The human evaluation mentioned, such as the evaluation of comments' labels, was performed by the authors of this paper and colleagues.
One potential negative outcome of this research is that it may reinforce stereotypes and biases that already exist. 
Additionally, the research may not generalize to all populations, and may not account for other factors such as age, culture, and education that could be influencing moral reasoning on social media.

\section*{Acknowledgments}

We thank the anonymous reviewers for their helpful comments. 
We thank the NSF (grant IIS-2116751) for partial support


\end{document}